\documentclass[aps,prl]{revtex4}
\newcommand{\beq}{\begin{equation}} \newcommand{\eeq}{\end{equation}}
\newcommand{\bea}{\begin{eqnarray}} \newcommand{\eea}{\end{eqnarray}}
\newcommand{\bear}{\begin{eqnarray*}} \newcommand{\eear}{\end{eqnarray*}}
\newcommand{\lb}{\label} 
\newcommand{\rf}[1]{(\ref{#1})}   
\usepackage[pdftex]{graphicx}
\usepackage{epstopdf}
\usepackage{amsmath}
\usepackage{amsfonts}
\usepackage{amssymb}
\usepackage{makeidx}

\begin{document}

\title {The Matrix Product Ansatz for integrable $U(1)^N$ models in Lunin-Maldacena backgrounds.}

\author{Matheus Jatkoske Lazo\footnote{lazo@smail.ufsm.br}} 
\address{Centro Tecnol\'ogico de Alegrete, UFSM/Unipampa, \\ Alegrete, RS, Brazil. \\ Programa de P\'os-gradua\c c\~ao em F\'\i sica, UFSM, 97111-900, \\ Santa Maria, RS, Brazil}

\begin{abstract}

We obtain through a Matrix Product Ansatz (MPA) the exact solution of the most general $N$-state spin chain with $U(1)^N$ symmetry and nearest neighbour interaction. In the case $N=6$ this model contain as a special case the integrable $SO(6)$ spin chain related to the one loop mixing matrix for anomalous dimensions in ${\cal N} = 4$ SYM, dual to type $IIB$ string theory in the generalised Lunin-Maldacena backgrounds. This MPA is construct by a map between scalar fields and abstract operators that satisfy an appropriate associative algebra. We analyses the Yang-Baxter equation in the $N=3$ sector and the consistence of the algebraic relations among the matrices defining the MPA and find a new class of exactly integrable model unknown up to now.



\end{abstract}

\maketitle

\noindent

\section{Introduction}

Since the discovery of the relation between the planar dilatation operator of the ${\cal{N}}=4$ Super Yang-Mills with an integrable $so(6)$ quantum spin chains \cite{MZ}, integrability has played a prominent role in the exploration of the Maldacena's correspondence between the ${\cal{N}}=4$ Super Yang-Mills with IIB string theory in $AdS_5\times S^5$ spaces \cite{Maldacena,GKP,Witten}. The study of the planar dilatation operator's integrability is very important because it not only enable us to test the Maldacena's correspondence as it is an generator of nontrivial integrable models. Exactly solvable models are of interest in both physics and mathematics since the pioneering work of Hans Bethe \cite{bethe}. The Bethe ansatz and its generalisations emerged over the years as a quite efficient and powerful tool for the exact solution of the eigenspectrum of a great variety of one dimensional quantum chains and two-dimensional transfer matrices (see, e.g., \cite{baxter,revkore,revessler,revschlo} for reviews). According to this ansatz the amplitudes of the eigenfunction are expressed by a nonlinear combination of properly defined plane waves. On the other hand, in the last two decades several different ansatz were introduced in the literature under the general name of matrix product ansatz (MPA). The first formulation was done for the description of the ground-state eigenfunction of some special non-integrable quantum chains, the so called valence-bond solid models \cite{affleck,arovas,fannes,klumper}. The MPA becomes also a successful tool for the exact calculation of the stationary probability distribution of some stochastic one dimensional systems \cite{derr2,derr1,alcritda}. An extension of this last MPA, called dynamical MPA was introduced in
\cite{schutz1,schutz2} and extended in \cite{popkov1}. This last ansatz gives the time-dependent probability distribution for some exact integrable systems. The MPA we are going to use in this paper, in order to solve the $U(1)^N$ quantum spin chain, was introduced in \cite{alclazo1,alclazo2,alclazo3,alclazo4}. This ansatz was applied with success in the evaluation of the spectra of several integrable quantum Hamiltonians \cite{alclazo1,alclazo2,alclazo3}, transfer matrices \cite{ice,lazo,alclazo5} and the time-evolution operator of stochastic systems \cite{alclazo4}. According to this ansatz, the amplitudes of the eigenfunctions are given in terms of a product of matrices where the matrices obey appropriated algebraic relations. In the case of the Bethe ansatz the spectral parameters and the amplitudes of the plane waves are fixed, apart from a normalisation constant, by the eigenvalue equation of the Hamiltonian or transfer matrix. On the other hand, in the MPA the eigenvalue equation fixes the commutation relations of the matrices defining the ansatz. In such case the spectrum of the Hamiltonian or transfer matrix, and the corresponding eigenfunctions, can be computed in a purely algebraic way.

In the present paper we study the most general quantum spin chain with $U(1)^N$ symmetry with nearest neighbour interaction thought a MPA in the generalised Lunin-Maldacena backgrounds (in this case we are only interested in zero momentum eigenstates, the more general case, including all momentum states, will be presented in \cite{lazo2}). This model contains as a special case the integrable $so(6)$ spin chain related to the one loop planar dilatation operator of the ${\cal{N}}=4$ Super Yang-Mills \cite{MZ}. We analyse the Yang-Baxter equation in the $N=3$ sector and the consistence of the algebraic relations among the matrices defining the ansatz and find that the solutions are separated in two class. In the first (class A) we obtain the models presented in \cite{FKM,BR,BC,BS}, in the context of one loop dilatation operator, as well as in condensed matter physics and stochastic models \cite{popkov1,alclazo2,alclazo4,perkshultz,sutherland,schlo}. In the second class (class B) we obtain the model presented in \cite{popkov1} for the stochastic problem of fully asymmetric diffusion of two kinds of particles. In this last sector we also find a new type of integrable model unknown up to now. Our solution generalises the previous results obtained in \rf{14} through the coordinate Bethe ansatz for $U(1)^N$ quantum spin chains. The analyses of the model for the full sector with $U(1)^N$ symmetry will be presented elsewhere \cite{lazo2}.


\section{The $U(1)^N$ spin chain in the generalised Lunin-Maldacena backgrounds}

The AdS/CFT conjecture relates operators, states, correlation functions and dynamics between the ${\cal{N}}=4$ Super Yang-Mills with IIB string theory in $AdS_5\times S^5$ spaces. One of the most important results of this conjecture predicts that the spectrum of scaling dimension $D$ of gauge invariant operators, in the conformal field theory, should coincide with the spectrum of energies $E$ of string states. Furthermore, this correspondence relates the weak coupling constant regime, in the gauge theory, with the strong coupling constant ones, in the string theory. Currently it is only possible to test the Maldacena's conjecture for a limited class of operators, and only in the case of large $\lambda$ values of the 't Hooft coupling constant. Operators where this conjecture can be tested are called BMN (Berenstein, Maldacena e Nastase) operators \cite{BMN}. These operators are local primary operators in $S^5$ with large $J$ charge, in the planar limit (large $\lambda$). These operator are given by the trace of a product of a large number the six scalar fields of the theory. The most general operator $O$ in this class is given by 
\beq
\lb{1}
O(\Psi,N)=\Psi^{\alpha_1...\alpha_L}_N{\mbox Tr}(\Phi_{\alpha_1}\cdots\Phi_{\alpha_L}),
\eeq
where $\alpha_j=1,...,N$, with $N=6$ and $j=1,...L$, and $L>>1$ is the number of fields. Instead of restrict our study to the case $N=6$, we will consider in this work $N$ general. The planar dilatation operator $D$, that is one of the generators of the conformal algebra and it give us the scaling dimensions of local operators, and consequently give us the energies of the string's states, is given by:
\beq
\lb{2}
D(g)= L +g^2 H +{\mbox O}\left(\frac{1}{N}\right),
\eeq    
where $g=\frac{g^2_{YM}N}{8{\pi}^2}$, and the operator $L$ counts the number of scalar fields on the operators ${O}$ \rf{1}. The operator $H$ is the one loop planar dilatation operator. This operator $H$ was identified \cite{MZ} with an exactly integrable uni dimensional $SO(6)$ quantum spin chain Hamiltonian. By this expansion in loops \rf{2} it can perturbativelly compare the string theory in the BMN limit to the field theory. On the other hand, this equivalence between the dilatation operator with integrable spin chain makes possible to use the powerful Bethe {\it
ansatz} to diagonalise the one-loop dilatation operator \cite{MZ}. By diagonalising this operator through the Bethe {\it ansatz}, it is obtained the scaling dimensions of any state of local operators in the conformal field theory in a purely algebraic way. The integrability also guarantees that these scaling dimensions can be exactly obtained for all gauge group, and in particular, for the $U(N)$ group \cite{BKS}. On the side of string theory, the integrable structures was discovered by the observation that Green Schwartz superstrings in $AdS_5\times S^5$ has an infinite set of conserved non-local charges \cite{BPR}. These results have made possible the comparison between gauge theories and string theories in the plane wave limit \cite{SV}, furthermore, the integrability enable us to make very precise tests for the correspondence AdS/CFT.

The most general one-loop dilatation operator is related with an integrable spin model with long-range couplings \cite{Beisert}. In the particular case of only two types of scalar fields on the BMN operators \rf{2} (case $N=2$), this integrable Hamiltonian reduces to the famous XXX model \cite{MZ}. On the other hand, the case of $N=3$ was studied by Frolov e Tseytlin \cite{FT} and it
corresponds to a spin-$1$ spin chain. The most general spin-$1$ model in condensed matter physics \cite{alclazo2} was obtained by us in a recent work, but it is not studied with the objective of classifying all possible solutions. In the context of the one-loop dilatation operator the solution, on class A presented here, was obtained in \cite{FKM}.

We consider here the most general $N$-state spin chain with nearest neighbour interaction, periodic boundary condition, zero momentum eigenstates \footnote{The periodic boundary condition of the spin chain and the zero momentum eigenstates are a consequence of the trace operation defining the
  BMN operators \rf{1}}, and $U(1)^N$ symmetry. The $N$ possible states configurations of a given site is related to the $N$ different types of scalar fields in \rf{1}. The $U(1)^N$ symmetry imply that the Hamiltonian describing the time evolution of this spin chain conserves the number of states of each type. By denoting the basis of states at a given site as $|\alpha\rangle$ ($\alpha=1,...,N$), the Hamiltonian in a periodic lattice with $L$ sites takes the form 
\beq
H=\sum_{j=1}^L \left(
\sum_{\alpha \ne \beta =1}^N
\Gamma_{\beta\;\alpha}^{\alpha\;\beta}E_j^{\beta\;\alpha}E_{j+1}^{\alpha\;\beta}+\sum_{\alpha
  ,\beta=1}^N \Gamma_{\alpha\;\beta}^{\alpha\;\beta}E_j^{\alpha\;\alpha}E_{j+1}^{\beta\;\beta}\right),
\lb{3}
\eeq
where $E^{\alpha\;\beta}$ are $N \times N$ Wyel matrices with elements $\left(E^{\alpha\;\beta}\right)_{i,j}=\delta_{\alpha,i}\delta_{\beta,j}$ ($\alpha,\beta=1,...,N$). While the first term in the right hand side of \rf{3} acts over neighbour sites exchanging its configuration $|\alpha\rangle \otimes|\beta\rangle \rightarrow |\beta\rangle \otimes |\alpha\rangle$ with rate $\Gamma_{\beta\;\alpha}^{\alpha\;\beta}$, the second one is a diagonal operator with weight $\Gamma_{\alpha\;\beta}^{\alpha\;\beta}$. The eigenfunction for \rf{3} can be construct as 
\beq
\lb{4}
|O(\Psi,N)\rangle=\sum_{\alpha_1,...,\alpha_L}^* \Psi^{\alpha_1...\alpha_L}_N|\alpha_1,...,\alpha_L\rangle
\;\;\; (\alpha_j=1,...,N),
\eeq
where the amplitudes $\Psi^{\alpha_1...\alpha_L}_N$ are related to the amplitudes in \rf{1}, for the operator $O$, and the symbol $(*)$ in the sum denotes the restriction to the sets $\{\alpha_1,...,\alpha_L\}$ with the same number $n_{\alpha}$ of spins in configuration $\alpha$.


\section{The MPA}

In order to formulate a MPA for the Hamiltonian \rf{3}, we make a one-to-one correspondence between configurations of spins (or scalar fields in the SYM) and products of abstract matrices. This matrix product is construct by making a correspondence between sites with spin configuration $\alpha=1,...,N$ and a matrix $A^{(\alpha)}$. Our MPA asserts that the components of the amplitude of the eigenfunction $\Psi^{\alpha_1...\alpha_L}_N$ in \rf{4} are obtained by associating them
to a products of these matrices $A^{(\alpha)}$. Actually $A^{(\alpha)}$ ($\alpha=1,...,N$) are abstract operators with an associative product. A well defined eigenfunction is obtained, apart from a normalisation factor, if all the amplitudes are related uniquely. Equivalently, in the subset of words (products of matrices) in the algebra containing $n_{\alpha}$ matrices $A^{(\alpha)}$ ($\alpha=1,...,N$, $n_1+\cdots n_N=L$) there exists only a single independent 
word. The relation between any two words gives the ratio between the corresponding amplitudes of the components of the eigenfunction $|O(\Psi,N)\rangle$. To formulated the ansatz we can choose any uniform operation on the matrix products that gives a non-zero scalar. For a quantum chains with periodic boundary conditions the trace operation is a convenient chose to produce this
scalar \footnote{the ansatz \rf{4} differs from the ansatz proposed in \cite{alclazo1,alclazo2,alclazo3} by an additional matrix $\Omega_{P}$ introduced in \cite{alclazo1,alclazo2,alclazo3} to include non-zero momentum eigenfunctions.}. The amplitudes $\Psi^{\alpha_1...\alpha_L}_N$ in \rf{4} takes the form:
\beq
\lb{5}
\Psi^{\alpha_1...\alpha_L}_N = {\mbox Tr} (A^{(\alpha_1)}A^{(\alpha_2)}\cdots A^{(\alpha_L)}) \;\;\; (\alpha_j=1,...,N).
\eeq

It is obvious that the $N$ states $|\alpha...\alpha\rangle$ ($\alpha=1,...,N$) are all eigenstates of the Hamiltonian \rf{3}. In the following we shall choose $|1...1\rangle$ as our reference state. The Hamiltonian \rf{3} when applied to the components of the eigenfunction \rf{4} where we do not have spins configurations $|\alpha\rangle$ ($\alpha=2,...,N$) at nearest neighbour sites give us the constraints, for the amplitudes $\Psi^{\alpha_1...\alpha_L}_N$ \rf{5},
\bea
\label{6}
&&\!\!\!\!\!\!\!\varepsilon_n {\mbox Tr} (A^{x_1-1}A^{(\alpha_1)} A^{x_2-x_1-1} \cdots
A^{(\alpha_{j-1})} A^{x_j-x_{j-1}-1}A^{(\alpha_j)} A^{x_{j+1}-x_j-1}A^{(\alpha_{j+1})}\cdots A^{(\alpha_n)} A^{L-x_n})\nonumber \\
&& \;\;\;\;\;\; =\sum_{j=1}^n \left[ \Gamma_{1 \;\; \alpha_j}^{\alpha_j \;1} {\mbox Tr} ( A^{x_1-1}A^{(\alpha_1)}A^{x_2-x_1-1} \cdots
A^{(\alpha_{j-1})}A^{x_j-x_{j-1}-2}A^{(\alpha_j)}A^{x_{j+1}-x_j}A^{(\alpha_{j+1})}\cdots A^{(\alpha_n)}A^{L-x_n})\right.\nonumber \\
&& \;\;\;\;\;\; +\left. \Gamma_{\alpha_j \; 1}^{1\;\; \alpha_j} {\mbox Tr} ( A^{x_1-1}A^{(\alpha_1)}E^{x_2-x_1-1} \cdots A^{(\alpha_{j-1})}A^{x_j-x_{j-1}}A^{(\alpha_j)}A^{x_{j+1}-x_j-2}A^{(\alpha_{j+1})}\cdots
A^{(\alpha_n)}A^{L-x_n}) \right]\nonumber \\
&&\;\;\;\;\;\;+[(L-2n)\Gamma_{1 \; 1}^{1\; 1} + \sum_{l=1}^n
\left(\Gamma_{\alpha_l \; 1}^{\alpha_l \; 1}+ \Gamma_{1\; \alpha_l}^{1\;
    \alpha_l}\right)]{\mbox Tr} (A^{x_1-1}A^{(\alpha_1)} A^{x_2-x_1-1} \cdots
A^{(\alpha_n)} A^{L-x_n}) \;\;\;\;\; (\alpha_j=2,...,N),
\eea
where $\varepsilon_n$ is the energy of the eigenfunction \rf{4}, $A\equiv A^{(1)}$, $n=n_2+n_3+\cdots +n_N$, and $x_1,...,x_n$ are the position in the spin chain where we have a state configuration $|\alpha\neq 1\rangle$. A convenient solution of this last equation is obtained by identifying the matrices $A^{(\alpha)}$ ($\alpha=2,...,N$) as composed by spectral-parameter-dependent matrices \cite{alclazo1,alclazo2,alclazo3}. The distinguibility of states configurations allows two types of solutions. The standard solution (class A) is obtained if each of the matrices $A^{(\alpha)}$ ($\alpha=2,...,N$) is composed of $n=n_2+n_3+\cdots +n_N$ spectral parameter dependent matrices \cite{alclazo2,alclazo4}. The second class of solutions (class B) is obtained if
matrices $A^{(\alpha)}$ ($\alpha=2,...,N$) with different $\alpha$ value are composed of by distinct sets of spectral parameters matrices \cite{alclazo2}.

{\bf Solutions of class A}

In this case, the matrices $A^{(\alpha)}$ ($\alpha=2,...,N$) can be written in terms of the matrix $A$ and $n=n_2+n_3+\cdots +n_N$ spectral parameter dependent matrices $A^{(\alpha)}_{k_j}$ \footnote{The most general relation $A^{(\alpha)}=\sum_{j=1}^n A^{a} A^{(\alpha)}_{k_j}A^{b}$
  could be used. However \rf{6} is more convenient since otherwise the $S$-matrix in \rf{12} will depend on $a$ an $b$}:
\beq
\lb{7}
A^{(\alpha)}=\sum_{j=1}^n A^{(\alpha)}_{k_j}A, \;\;\;\; (\alpha=2,...,N),
\eeq
where the matrices $A^{(\alpha)}_{k_j}$ ($\alpha=2,...,N$) satisfy the following commutation relations with the matrix $A$:
\beq
\lb{8}
A^{(\alpha)}_{k_j}A=g(\alpha)e^{ik_j}AA^{(\alpha)}_{k_j},  \;\;\; (\alpha=2,...,N),  \;\;\; (j=1,...,n),
\eeq
the parameters $k_j$ ($j=1,...,n$) are in general complex numbers unknown {\it a priori}, and $g(\alpha)$ is a constant. The energy $\varepsilon_n$ is obtained by inserting \rf{7} in \rf{6}, by using \rf{8} and imposing that $\varepsilon_n$ is a symmetric function on the spectral parameters:
\beq
\lb{9}
\varepsilon_n = \sum_{j=1}^n \left(\Gamma_{1\;2}^{2\;1}e^{ik_j}+ \Gamma_{2\;1}^{1\;2}e^{-ik_j}\right) + 
\sum_{\alpha=2}^N n_{\alpha}\left(\Gamma_{1\;\alpha}^{1\;\alpha}+\Gamma_{\alpha\;1}^{\alpha\;1}\right)+(L-2n)\Gamma_{1\;1}^{1\;1},
\eeq
where we need to impose
\beq
\lb{10}
g(\alpha)=\frac{\Gamma_{1\;2}^{2\;1}}{\Gamma_{1\;\alpha}^{\alpha\;1}}=\frac{\Gamma_{\alpha\;1}^{1\;\alpha}}{\Gamma_{2\;1}^{1\;2}} \;\;\; (g(2)=1, \;\;\; \alpha=3,...,N).
\eeq

The relations coming from the eigenvalue equation for configurations where we have two spins configurations $|\alpha\rangle$ ($\alpha=2,...,N$) at nearest neighbour sites are given by
\bea
\lb{11}
&&\sum_{j,l=2}^n\left[\Gamma_{2\;1}^{1\;2}+\Gamma_{1\;2}^{2\;1}e^{i(k_j+k_l)}+(\Gamma_{\alpha\;1}^{\alpha\;1}+\Gamma_{1\;\alpha}^{1\;\alpha}-\Gamma_{1\;1}^{1\;1}-\Gamma_{\alpha\;\alpha}^{\alpha\;\alpha})e^{ik_j} \right]A^{(\alpha)}_{k_j}A^{(\alpha)}_{k_l}=0,  \\
&&\sum_{j,l=2}^n\left[\Gamma_{2\;1}^{1\;2}+\Gamma_{1\;2}^{2\;1}e^{i(k_j+k_l)}+(\Gamma_{\alpha\;1}^{\alpha\;1}+\Gamma_{1\;\beta}^{1\;\beta}-\Gamma_{1\;1}^{1\;1}-\Gamma_{\alpha\;\beta}^{\alpha\;\beta})e^{ik_j}
\right]A^{(\alpha)}_{k_j}A^{(\beta)}_{k_l}=\sum_{j,l=2}^n\frac{g(\beta)}{g(\alpha)}\Gamma_{\alpha\;\beta}^{\beta\;\alpha}e^{ik_l}A^{(\beta)}_{k_j}A^{(\alpha)}_{k_l}\;\;\;
(\alpha \neq \beta), \nonumber
\eea
where we have used \rf{5}, \rf{7}-\rf{10}. The relations \rf{11} fix the algebraic relations among the matrices $A^{(\alpha)}_{k_j}$ $(\alpha=2,...,N)$:
\beq
\lb{12}
A^{(\alpha)}_{k_j}A^{(\beta)}_{k_l}=\sum_{\alpha',\beta'=2}^N S^{\alpha \; \beta}_{\beta' \; \alpha'}(k_j,k_l)A^{(\alpha')}_{k_l}A^{(\beta' )}_{k_j},  \;\;\;A^{(\alpha)}_{k_j}A^{(\beta)}_{k_j}=0 \;\;\; (l \neq j =1,...,n),
\eeq
where the nonzero structural constants $S^{\alpha \; \beta}_{\beta' \;
  \alpha'}(k_j,k_l)$ are given by:
\bea
\lb{13}
&&S^{\alpha \; \alpha}_{\alpha \; \alpha}(k_j,k_l)=-\frac{\Gamma_{2\;1}^{1\;2}+\Gamma_{1\;2}^{2\;1}e^{i(k_j+k_l)}+(\Gamma_{\alpha\;1}^{\alpha\;1}+\Gamma_{1\;\alpha}^{1\;\alpha}-\Gamma_{1\;1}^{1\;1}-\Gamma_{\alpha\;\alpha}^{\alpha\;\alpha})e^{ik_l}}{\Gamma_{2\;1}^{1\;2}+\Gamma_{1\;2}^{2\;1}e^{i(k_j+k_l)}+(\Gamma_{\alpha\;1}^{\alpha\;1}+\Gamma_{1\;\alpha}^{1\;\alpha}-\Gamma_{1\;1}^{1\;1}-\Gamma_{\alpha\;\alpha}^{\alpha\;\alpha})e^{ik_j}}, \nonumber \\
&&S^{\alpha \; \beta}_{\beta \;
  \alpha}(k_j,k_l)=-\frac{C_{\alpha,\beta}(k_l,k_j)C_{\beta,\alpha}(k_j,k_l)-\Gamma_{\alpha\;\beta}^{\beta\;\alpha}\Gamma_{\beta\;\alpha}^{\alpha\;\beta}e^{i(k_j+k_l)}}{C_{\alpha,\beta}(k_j,k_l)C_{\beta,\alpha}(k_j,k_l)-\Gamma_{\alpha\;\beta}^{\beta\;\alpha}\Gamma_{\beta\;\alpha}^{\alpha\;\beta}e^{2ik_j}}, \;\;\; (\alpha \neq \beta =2,...,N) \\
&&S^{\alpha \; \beta}_{\alpha \;
  \beta}(k_j,k_l)=-\frac{g(\beta)}{g(\alpha)}\frac{\Gamma_{\alpha\;\beta}^{\beta\;\alpha}(C_{\beta,\alpha}(k_l,k_j)e^{ik_j}-C_{\beta,\alpha}(k_j,k_l)e^{ik_l})}{C_{\alpha,\beta}(k_j,k_l)C_{\beta,\alpha}(k_j,k_l)-\Gamma_{\alpha\;\beta}^{\beta\;\alpha}\Gamma_{\beta\;\alpha}^{\alpha\;\beta}e^{2ik_j}}, \nonumber \\
&&C_{\alpha,\beta}(k_j,k_l)=\Gamma_{2\;1}^{1\;2}+\Gamma_{1\;2}^{2\;1}e^{i(k_j+k_l)}+(\Gamma_{\alpha\;1}^{\alpha\;1}+\Gamma_{1\;\beta}^{1\;\beta}-\Gamma_{1\;1}^{1\;1}-\Gamma_{\alpha\;\beta}^{\alpha\;\beta})e^{ik_j}. \nonumber
\eea
Relations \rf{8} and \rf{12} define completely the algebra whose structural constants are the S-matrix given by \rf{13}. It is important to mention that in the sector $N=2$ (only two types of scalar fields $\phi_{1}$ and $\phi_{2}$, for example) the Hamiltonian \rf{3} reduces to the well known asymmetric XXZ model \cite{yangyang}. In this particular case the $S$-matrix is a diagonal
matrix were $S^{2 \; 2}_{2 \; 2}(k_j,k_l)$ in \rf{13} is the only nonzero element. As a consequence, the algebra defined by \rf{12} is associative for $N=2$. For $N$ general the $S$-matrix is not diagonal and the algebra \rf{12} is not associative for arbitrary values of
$\Gamma_{\beta\;\alpha}^{\alpha\;\beta}$ and $\Gamma_{\alpha\;\beta}^{\alpha\;\beta}$. For arbitrary amplitudes we have in our matrix product ansatz \rf{5} a product of $n$ matrices
$A^{(\alpha)}_{k_j}$. Our ansatz will be valid only if the relations \rf{12} provide a unique relation among these products, otherwise the eigenfunction \rf{4} is not properly defined. This means, for example, that the products $\cdots A_{k_1}^{(\alpha)}A_{k_2}^{(\beta)}A_{k_3}^{(\gamma)}\cdots$ and $\cdots A_{k_3}^{(\gamma)}A_{k_2}^{(\beta)}A_{k_1}^{(\alpha)}\cdots$ should be
uniquely related. Since we can relate then either by performing the commutations in the order $\alpha \beta \gamma \rightarrow \beta \alpha \gamma \rightarrow \beta \gamma \alpha \rightarrow \gamma \beta \alpha$ or $\alpha \beta \gamma \rightarrow \alpha \gamma \beta \rightarrow \gamma \alpha \beta \rightarrow \gamma \beta \alpha$, the structure constants $S_{\gamma\;\gamma'}^{\alpha\;\alpha'}$ of the algebraic relations \rf{12} should satisfy
\beq
\sum_{\gamma,\gamma',\gamma''=2}^N S_{\gamma\;\gamma'}^{\alpha\;\alpha'}(k_1,k_2)S_{\beta\;\gamma''}^{\gamma\;\alpha''}(k_1,k_3)S_{\beta'\;\beta''}^{\gamma'\;\gamma''}(k_2,k_3)=
\sum_{\gamma,\gamma',\gamma''=2}^N S_{\gamma'\;\gamma''}^{\alpha'\;\alpha''}(k_2,k_3)S_{\gamma\;\beta''}^{\alpha\;\gamma''}(k_1,k_3)S_{\beta\;\beta'}^{\gamma\;\gamma'}(k_1,k_2),
\lb{14}
\eeq
for $\alpha, \alpha', \alpha'', \beta, \beta', \beta''=2,...,6$.  This last constraint is just the Yang-Baxter relation \cite{baxter,Yang2} of the $S$-matrix defined in \rf{13}.  Actually the condition \rf{14} is enough to ensure that any matrix product of spectral matrices $\{A_{k_j}^{(\alpha)}\}$ is uniquely related and it implies the associativity of the algebra of the matrices $\{A_{k_j}^{(\alpha)}\}$ . In this case the relations coming from the eigenvalue equation for configurations where we have more than two spins configurations $|\alpha\rangle$ ($\alpha=2,...,N$) at nearest neighbour sites are automatically satisfied by the algebraic relations \rf{8} and \rf{12} among the matrices defining the ansatz, no new constraints appear from these relations. On the other hand, the Yang-Baxter relation \rf{14} produces strong constraints in the allowed couplings $\Gamma_{\beta\;\alpha}^{\alpha\;\beta}$ and $\Gamma_{\alpha\;\beta}^{\alpha\;\beta}$ of the Hamiltonian \rf{3}. In the sector $N=3$ (tree types of scalar fields $\phi_{1}$, $\phi_{2}$ and
$\phi_{3}$, for example) the Hamiltonian \rf{3} reduces to an asymmetric spin-$1$ quantum Hamiltonian with $U(1)^3$ symmetry. The spin-$1$ model with $g(2)=g(3)=1$ in \rf{10} and $U(1)^3$ symmetry was obtained by us in a recent work \cite{alclazo2} in the context of condensed matter physics, but it is not studied with the objective of classifying all possible solutions of the
Yang-Baxter equation \rf{14}. In the generalised Lunin-Maldacena backgrounds, the solutions of \rf{14} was classified in \cite{FKM} for hermitian Hamiltonians \rf{3}. In the present paper we obtain the most general solution of \rf{14}. These solutions generalises the previous results obtained in \cite{alclazo2,FKM} and can be obtained by a systematic investigation with the aid of Maple by purely analytical means. We find the following types of solutions:
\beq
\lb{14a}
\begin{split}
&{\mbox (A.1)}\;\;\;\;\; \Gamma_{\alpha\;1}^{1\;\alpha}\Gamma_{1\; \alpha}^{\alpha\;1}=\Gamma_{\beta\;1}^{1\;\beta}\Gamma_{1\;\beta}^{\beta \; 1}=\Gamma_{\alpha\;\beta}^{\beta\;\alpha}\Gamma_{\beta\;\alpha}^{\alpha\;\beta}=t_{\alpha\beta1}t_{\beta\alpha1}, \;\;\;\;\; t_{\alpha\alpha1}=t_{\beta\beta1}=0,\\
&{\mbox (A.2)}\;\;\;\;\; \Gamma_{\alpha\;1}^{1\;\alpha}\Gamma_{1\; \alpha}^{\alpha\;1}=\Gamma_{\beta\;1}^{1\;\beta}\Gamma_{1\;\beta}^{\beta \; 1}=\Gamma_{\alpha\;\beta}^{\beta\;\alpha}\Gamma_{\beta\;\alpha}^{\alpha\;\beta}=t_{\alpha\beta1}t_{\beta\alpha1}, \;\;\;\;\; t_{\alpha\alpha1}=t_{\alpha\alpha\beta}=0,\\
&{\mbox (A.3)}\;\;\;\;\; \Gamma_{\alpha\;1}^{1\;\alpha}\Gamma_{1\; \alpha}^{\alpha\;1}=\Gamma_{\beta\;1}^{1\;\beta}\Gamma_{1\;\beta}^{\beta \; 1}=\Gamma_{\alpha\;\beta}^{\beta\;\alpha}\Gamma_{\beta\;\alpha}^{\alpha\;\beta}=t_{\alpha\beta1}t_{\beta\alpha1}, \;\;\;\;\; t_{1\beta\alpha}=t_{\alpha1\beta}=t_{\beta\alpha1},\\
&{\mbox (A.4)}\;\;\;\;\; \Gamma_{\alpha\;1}^{1\;\alpha}\Gamma_{1\; \alpha}^{\alpha\;1}=\Gamma_{\beta\;1}^{1\;\beta}\Gamma_{1\;\beta}^{\beta \; 1}, \;\;\;\;\; \Gamma_{\alpha\;\beta}^{\beta\;\alpha}=\Gamma_{\beta\;\alpha}^{\alpha\;\beta}=0, \;\;\;\;\; t_{\alpha\beta1}=t_{\beta\alpha1}=t_{\beta\beta 1}=t_{\alpha\alpha 1},\\
&{\mbox (A.5)}\;\;\;\;\; \Gamma_{1\; \alpha}^{\alpha\;1}=g\Gamma_{1\;\beta}^{\beta\; 1},\;\;\;\;\; \Gamma_{\alpha\;1}^{1\;\alpha}=\Gamma_{\beta\;1}^{1\;\beta}=0,\;\;\;\;\; \Gamma_{\beta\;\alpha}^{\alpha\;\beta}=0, \;\;\;\;\; t_{\alpha\beta1}=t_{\alpha\alpha1},\\
&{\mbox (A.6)}\;\;\;\;\; \Gamma_{1\; \alpha}^{\alpha\;1}=g\Gamma_{1\;\beta}^{\beta\; 1},\;\;\;\;\; \Gamma_{\alpha\;1}^{1\;\alpha}=\Gamma_{\beta\;1}^{1\;\beta}=0,\;\;\;\;\; \Gamma_{\beta\;\alpha}^{\alpha\;\beta}=0, \;\;\;\;\; t_{\beta\alpha1}=t_{\alpha\alpha1},\\
&{\mbox (A.7)}\;\;\;\;\; \Gamma_{\alpha\; 1}^{1\; \alpha}=g\Gamma_{\beta\; 1}^{1\; \beta},\;\;\;\;\; \Gamma_{1\; \alpha}^{\alpha\; 1}=\Gamma_{1\; \beta}^{\beta\; 1}=0,\;\;\;\;\; \Gamma_{\beta\;\alpha}^{\alpha\;\beta}=0, \;\;\;\;\; t_{\alpha\beta1}=t_{\beta\beta1},\\
&{\mbox (A.8)}\;\;\;\;\; \Gamma_{\alpha\; 1}^{1\; \alpha}=g\Gamma_{\beta\;1}^{1\;\beta},\;\;\;\;\; \Gamma_{1\;\alpha}^{\alpha\;1}=\Gamma_{1\;\beta}^{\beta\;1}=0,\;\;\;\;\; \Gamma_{\beta\;\alpha}^{\alpha\;\beta}=0, \;\;\;\;\; t_{\beta\alpha1}=t_{\beta\beta1},\\
&{\mbox (A.9)}\;\;\;\;\; \Gamma_{\alpha\;1}^{1\;\alpha}=\Gamma_{1\; \alpha}^{\alpha\;1}=\Gamma_{\beta\;1}^{1\;\beta}=\Gamma_{1\;\beta}^{\beta \; 1}=0.
\end{split}
\eeq
where $\alpha,\beta=2,3$ with $\alpha \neq \beta$ and $t_{\alpha'\beta'\gamma}=\Gamma_{\alpha'\;\beta'}^{\alpha'\;\beta'}+\Gamma_{\gamma\;\gamma}^{\gamma\;\gamma}-\Gamma_{\alpha'\;\gamma}^{\alpha'\;\gamma}-\Gamma_{\gamma\;\beta'}^{\gamma\;\beta'}$ ($\alpha',\beta',\gamma=1,2,3$). The first four models A.1-A.4 and A.9 contain as a special case (for hermitian hamiltonians \rf{3}) the solutions obtained in \cite{FKM,BR,BC,BS}, in the context of one loop dilatation operator, as well as in condensed matter physics and stochastic models \cite{popkov1,alclazo2,alclazo4,perkshultz,sutherland,schlo}. The models A.5-A.8 generalises the model presented in \cite{popkov1} for the stochastic problem of fully asymmetric diffusion of two kinds of particles.

In order to complete our solutions through the MPA \rf{5} we should fix the spectral parameters, or momenta, $k_1,\ldots, k_n$. Using the algebraic relations \rf{7} and \rf{8} an arbitrary amplitude is proportional to $\mbox{Tr} \left[ A_{k_1}^{(\alpha_1)}\cdots  A_{k_n}^{(\alpha_n)} A^L\right]$. The cyclic property of the trace and the commutation relations \rf{8} and \rf{12} give us
\beq
\mbox{Tr} \left[ A_{k_1}^{(\alpha_1)}\cdots  A_{k_n}^{(\alpha_n)}A^L \right] = e^{ik_jL} \sum_{\alpha_1',\ldots,\alpha_n'=2}^N \langle
\alpha_1,\ldots,\alpha_n|{\cal{T}}^{(n)}|\alpha_1',\ldots,\alpha_n'\rangle \mbox{Tr}
\left[ A_{k_1}^{(\alpha_1')}\cdots  \alpha_{k_n}^{(\alpha_n')}E^L \right],
\lb{15}
\eeq
where we have used the identity (see \rf{13})
\beq
\sum_{\alpha_j'',\alpha_{j+1}''}S_{\alpha_j'\;\alpha_j''}^{\alpha_j\;\alpha_{j+1}''}(k_j,k_j)=-1,
\lb{16}
\eeq
and
\beq \langle \alpha_1,\ldots,\alpha_n|{\cal{T}}^{(n)}|\alpha_1',\ldots,\alpha_n'\rangle = \sum_{\alpha_1'',\ldots,\alpha_n''} \left\{ S_{\alpha_1'\;\alpha_1''}^{\alpha_1\;\alpha_2''}(k_1,k_j)\cdots S_{\alpha_j'\;\alpha_j''}^{\alpha_j\;\alpha_{j+1}''}(k_j,k_j)\cdots S_{\alpha_n'\;\alpha_n''}^{\alpha_n\;\alpha_1''}(k_n,k_j)\phi(\alpha_1'') \right\}
\lb{17}
\eeq
where $\phi(\alpha_1'')=g(\alpha_1'')^L$, is a $(N-1)^n \times (N-1)^n$-dimensional transfer matrix of an inhomogeneous vertex model (inhomogeneities $\{k_l \}$) with Boltzmann weights given by \rf{13}. The model is defined on a cylinder of perimeter $n$ with a seam along its axis producing the twisted boundary condition
\beq
\lb{17b}
S_{\alpha_n'\;\alpha_n''}^{\alpha_n\;\alpha_{n+1}''}(k_n,k_j)=S_{\alpha_n'\;\alpha_n''}^{\alpha_n\;\alpha_1''}(k_n,k_j)\phi(\alpha_1'').
\eeq
Finally relation \rf{15} with \rf{17b} give us the constraints for the spectral parameters:
\beq
e^{-ik_jL}=\Lambda^{(n)}(k_j,\{k_l \}) \;\;\; (j=1,\ldots,n), 
\lb{18}
\eeq
where $\Lambda^{(n)}(k_j,\{k_l \})$ are the eigenvalues of the transfer matrix \rf{17}. The condition \rf{18} leads to the problem of evaluation the eigenvalues of the inhomogeneous transfer matrix \rf{17}. This can be done through the algebraic Bethe  ansatz \cite{kulish} or the coordinate Bethe
ansatz (see \cite{alcbar4} and \cite{bjp4} for example).

{\bf Solutions of class B}

In this case, the matrices $A^{(\alpha)}$ ($\alpha=2,...,N$) with different $\alpha$ value are composed of by distinct sets of spectral parameters matrices. For simplicity we are going to consider that there are two sets of spectral parameters. For a given $\delta=2,...,N$, we have $n_a=n_2+\cdots +n_{\delta}$ parameters $k_j^{(a)}$ ($j=1,...,n_a$) and $n_b=n_{\delta+1}+\cdots +n_N$ parameters $k_j^{(b)}$ ($j=1,...,n_b$). In this case the matrices $A^{(\alpha)}$ can be written as:
\beq
\lb{19}
A^{(\alpha)}=\sum_{j=1}^{n_a} A^{(\alpha)}_{k_j^{(a)}}A, \;\;\;\; (\alpha=2,...,\delta), \;\;\;\; A^{(\alpha)}=\sum_{j=1}^{n_b} A^{(\alpha)}_{k_j^{(b)}}A, \;\;\;\; (\alpha=\delta+1,...,N),
\eeq
where the matrices $A^{(\alpha)}_{k_j}$ ($\alpha=2,...,N$) satisfy the following
commutation relations with the matrix $A$:
\beq
\lb{20}
A^{(\alpha)}_{k_j^{(\alpha)}}A=g(\alpha)e^{ik_j^{(\alpha)}}AA^{(\alpha)}_{k_j^{(\alpha)}},  \;\;\; (\alpha=2,...,N),
\eeq
with $k_j^{(\alpha)}=k_j^{(a)}$ for $\alpha \leq \delta$ and $k_j^{(\alpha)}=k_j^{(b)}$ for $\alpha > \delta$. The energy $\varepsilon_n$ is obtained by inserting \rf{19} in \rf{6}, by using \rf{20} and imposing that $\varepsilon_n$ is a symmetric function on each sets of spectral parameters:
\beq
\lb{21}
\varepsilon_n = \sum_{j=1}^{n_a} \left(\Gamma_{1\;2}^{2\;1}e^{ik_j^{(a)}}+ \Gamma_{2\;1}^{1\;2}e^{-ik_j^{(a)}}\right) + \sum_{j=1}^{n_b} \left(\Gamma_{1\;\delta+1}^{\delta+1\;1}e^{ik_j^{(b)}}+ \Gamma_{\delta+1\;1}^{1\;\delta+1}e^{-ik_j^{(b)}}\right)+
\sum_{\alpha=2}^N n_{\alpha}\left(\Gamma_{1\;\alpha}^{1\;\alpha}+\Gamma_{\alpha\;1}^{\alpha\;1}\right)+(L-2n)\Gamma_{1\;1}^{1\;1},
\eeq
where we need to impose
\beq
\lb{22}
g(\alpha)=\frac{\Gamma_{1\;2}^{2\;1}}{\Gamma_{1\;\alpha}^{\alpha\;1}}=\frac{\Gamma_{\alpha\;1}^{1\;\alpha}}{\Gamma_{2\;1}^{1\;2}} \;\;\; (g(2)=1, \;\;\;\alpha=3,...,\delta),\;\;\;\;\; g(\alpha)=\frac{\Gamma_{1\;\delta+1}^{\delta+1\;1}}{\Gamma_{1\;\alpha}^{\alpha\;1}}=\frac{\Gamma_{\alpha\;1}^{1\;\alpha}}{\Gamma_{\delta+1\;1}^{1\;\delta+1}} \;\;\; (g(\delta+1)=1, \;\;\; \alpha=\delta+2,...,N).
\eeq

As in case A, the algebraic relations among the matrices $A^{(\alpha)}_{k_j}$ $(\alpha=2,...,N)$ are fixed by the eigenvalue equation for the Hamiltonian \rf{3} when applied to the components of the eigenfunction \rf{4} where we have two spins configurations $|\alpha\rangle$ ($\alpha=2,...,N$) at nearest neighbour sites, but now we have two situations. When we have spins configuration $\alpha,\beta\leq \delta$ or $\alpha,\beta > \delta$ located at closest positions, and when we have $\alpha\leq \delta $ ($\beta\leq \delta $) and $\beta > \delta $ ($\alpha >\delta $). In this last case, in order to satisfy the equation \rf{4} we need to impose the following constraints:
\beq
\lb{23}
\begin{split}
&{\mbox (B.1)}\;\;\;\;\; \Gamma_{\alpha\;1}^{1\;\alpha}=\Gamma_{1\;\beta}^{\beta\;1}=0, \;\;\;\;\; \Gamma_{\alpha\;\beta}^{\beta\;\alpha}=0,\;\;\;\;\; t_{\alpha\beta1}=0\\
&{\mbox (B.2)}\;\;\;\;\; \Gamma_{\beta\;1}^{1\;\beta}=\Gamma_{1\;\beta}^{\beta\;1}=0, \;\;\;\;\; \Gamma_{\alpha\;\beta}^{\beta\;\alpha}\Gamma_{\beta\;\alpha}^{\alpha\;\beta}=\Gamma_{\alpha\;1}^{1\;\alpha}\Gamma_{1\;\alpha}^{\alpha\;1},\;\;\;\;\; t_{\alpha\beta1}=t_{\beta\alpha1}=t_{\beta\beta1}=0,
\end{split}
\eeq
where $\alpha\leq \delta $ ($\beta\leq \delta $) and $\beta > \delta $ ($\alpha >\delta $), we also obtain the structural constants:
\beq
\lb{24}
\begin{split}
&{\mbox (B.1)}\;\;\;\;\; S^{\alpha \; \beta}_{\alpha \; \beta}(k_j^{(\alpha)},k_l^{(\beta)})=\frac{1}{S^{\beta \; \alpha}_{\beta \; \alpha}(k_l^{(\beta)},k_j^{(\alpha)})}=\frac{g(\beta)}{g(\alpha)}\frac{\Gamma_{\alpha\;\beta}^{\beta\;\alpha}e^{ik_j^{(\alpha)}}}{\Gamma_{2\;1}^{1\;2}+\Gamma_{1\;\delta+1}^{\delta+1\;1}e^{i(k_j^{(\alpha)}+k_l^{(\beta)})}+(\Gamma_{\alpha\;1}^{\alpha\;1}+\Gamma_{1\;\beta}^{1\;\beta}-\Gamma_{1\;1}^{1\;1}-\Gamma_{\alpha\;\beta}^{\alpha\;\beta})e^{ik_j^{(\alpha)}}},  \\
&{\mbox (B.2)}\;\;\;\;\; S^{\alpha \; \beta}_{\alpha \; \beta}(k_j^{(\alpha)},k_l^{(\beta)})=\frac{1}{S^{\beta \; \alpha}_{\beta \; \alpha}(k_l^{(\beta)},k_j^{(\alpha)})}=\frac{g(\beta)}{g(\alpha)}\frac{\Gamma_{\alpha\;\beta}^{\beta\;\alpha}}{\Gamma_{2\;1}^{1\;2}}e^{ik_j^{(\beta)}},
\end{split}
\eeq
where we have written for simplicity only the case $\alpha\leq \delta$ and $\beta >\delta$. On the other hand, when $\alpha,\beta\leq \delta$ or $\alpha,\beta > \delta $, we have a similar situation as that considered in the case A and we obtain algebraic relations with $S^{\alpha \; \beta}_{\beta' \; \alpha'}(k_j^{(\alpha)},k_l^{(\alpha)})$ as in \rf{12} and coupling constants given by \rf{23}. The only difference occur when we have $\alpha=\beta >\delta$ where
\beq
\lb{25}
{\mbox (B.2)} \;\;\;\;\; S^{\beta \; \beta}_{\beta \; \beta}(k_j^{(b)},k_l^{(b)})=1.
\eeq
In the sector $N=3$ the model B.1 generalises the stochastic problem of fully asymmetric diffusion of two kinds of particles, whose exact integrability was obtained in \cite{popkov1} through the dynamical matrix product ansatz. The case B.2 is a new type of integrable model unknown up to now. It is important to mention the B.2 type models can be made hermitian and it is related to the A.4 model by an interchange of labels $1$ and $\alpha$, however since its exactly solutions describe different sectors of spin configurations, they are different physical system.  

Finally, as in case A, the momenta $k_j^{(a)}$ ($j=1,...,n_a$) and $k_j^{(b)}$ ($j=1,...,n_b$) are fixed by the cyclic property of the trace in \rf{5}. We have for both B.1 and B.2:
\beq
\lb{26}
e^{-ik_j^{(a)}L}=\Lambda^{(n_a)}(k_j^{(a)},\{k_l^{(a)} \}) \;\;\;\;\; \mbox{and} \;\;\;\;\; e^{-ik_j^{(b)}L}=\Lambda^{(n_b)}(k_j^{(b)},\{k_l^{(b)} \}),
\eeq
where $\Lambda^{(n_a)}(k_j^{(a)},\{k_l^{(a)} \})$ and $\Lambda^{(n_b)}(k_j^{(b)},\{k_l^{(b)} \})$ are the eigenvalue of the transfer matrix defined in \rf{17} with $\phi(\alpha_1'')=g(\alpha_1'')^L\prod_{l=1}^{n_b}S^{\alpha_1'' \; \beta_l}_{\alpha_1'' \; \beta_l}(k_j^{(a)},k_l^{(b)})$ and $\phi(\beta_1'')=g(\beta_1'')^L\prod_{l=1}^{n_a}S^{\beta_1'' \; \alpha_l}_{\beta_1'' \; \alpha_l}(k_j^{(b)},k_l^{(a)})$, respectively, and where $S^{\alpha' \; \beta'}_{\beta'' \; \alpha''}(k_j^{(\alpha')},k_l^{(\beta')})$ is given by \rf{13}, \rf{24} and \rf{25} with \rf{23}.

\section{Discussion and conclusions}

We solve through a MPA the most general $N$-state spin chain with $U(1)^N$ symmetry and nearest neighbour interaction in Lunin-Maldacena backgrounds. According to this ansatz, the amplitudes of the eigenfunctions are given in terms of a product of matrices where the matrices obey appropriated algebraic relations. This MPA is constructed by making a one-to-one correspondence between configurations of spins (or scalar fields in the SYM) and products of abstract matrices. Although in Lunin-Maldacena backgrounds all eigenstates have zero momentum, we can generalise the MPA including all momentum states \cite{lazo2}. We analyses the Yang-Baxter equation in the $N=3$ sector and the consistence of the algebraic relations among the matrices defining the MPA and find a new class of exactly integrable model unknown up to now (model B.2). This new model is a consequence of the distinguibility of states configurations due to the $U(1)^N$ symmetry. The study of this new model can be of interest in the context of both AdS/CFT and stochastic process. MPA. On the other hand, it will be interesting to see if it is possible to formulate a MPA for the most general one-loop dilatation operator related with an integrable spin model with long-range couplings \cite{Beisert}. Another quite interesting problem for the future concerns the formulation of the MPA for the case where we have open boundary conditions, as well as for quantum chains with no global conservation law such as the XYZ model, the 8-vertex model or the case where the quantum chains are defined on open lattices with non-diagonal boundary fields.

\section{Acknowledgements}

I am grateful to F. C. Alcaraz for his discussions and P. A. de Castro for his reading of the paper. This work has been supported by CAPES and FAPESP (Brazilian agencies).

\end{document}